# Localized electronic states induced by defects and possible origins of ferroelectricity in strontium titanate thin films

Y. S. Kim,[1] J. Kim,[2] S. J. Moon,[1] W. S. Choi,[1] Y. J. Chang,[1] J.-G. Yoon,[3] J. Yu,[2] J.-S. Chung,[4] and T. W. Noh[1,a)]

[1]*ReCOE & FPRD, Department of Physics and Astronomy, Seoul National University, Seoul 151-747, Korea*
[2]*CSCMR & FPRD, Department of Physics and Astronomy, Seoul National University, Seoul 151-747, Korea*
[3]*Department of Physics, University of Suwon, Kyunggi-do 445-743, Korea*
[4]*Department of Physics, Soongsil University, Kyunggi-do 156-743, Korea*

Several oxygen vacancy defect configurations have been investigated as possible origins of reported room-temperature ferroelectricity in strontium titanate (STO) thin films [Appl. Phys. Letts. **91**, 042908 (2007)]. First-principles calculations revealed that the Sr–O–O vacancy complexes create deep localized states in the band gap of $SrTiO_3$ without affecting its insulating property. These results are in agreement with electronic structural changes determined from optical transmission and X-ray absorption measurements. This work opens the way to exploiting oxygen vacancies and their complexes as a source of ferroelectricity in perovskite oxide thin films, including STO.

$SrTiO_3$ is a well-known incipient ferroelectric (FE) material.[1] As quantum fluctuation and antiferrodistortion compete with FE instability, $SrTiO_3$ remains in paraelectric phase even at low temperatures. Many researchers have used the strain engineering concept based on a lattice mismatch with different substrates to improve the FE properties.[2,3] We recently successfully fabricated strontium titanate (STO) thin films that exhibited room-temperature ferroelectricity.[4] We introduced vacancy defects in the STO films by adjusting the growth conditions. Although we attributed the occurrence of ferroelectricity to the defects in the STO films,[4] it was unclear what types of defect play important roles in this ferroelectricity.

In this paper, we describe the investigation of possible defects and their roles in FE STO films. From optical spectroscopy, we found that the occurrence of ferroelectricity in STO films has a strong correlation with the appearance of an absorption peak at around 1.3 eV. By performing first-principles calculations, we investigated changes in the electronic structure due to several complexes of Sr and O vacancies. We found that the Sr–O–O vacancy complex is the most probable candidate for defect dipoles that provide a localized state in the band gap, resulting in the optical absorption around 1.3 eV and ferroelectricity.

For experimental investigation of the electronic structures of FE STO films, we grew 100-nm-thick STO films on both sides of polished $SrTiO_3$ (001) substrates for optical measurements using the same deposition conditions as previously reported.[4] Note that the sample was not transparent when using conductive $SrRuO_3$ (SRO) electrode layers, as in our previous research. All STO films were deposited under the same deposition conditions, except for oxygen partial pressure ($P_{O2}$), which ranged from $2\times10^{-2}$ to $1\times10^{-6}$ Torr.

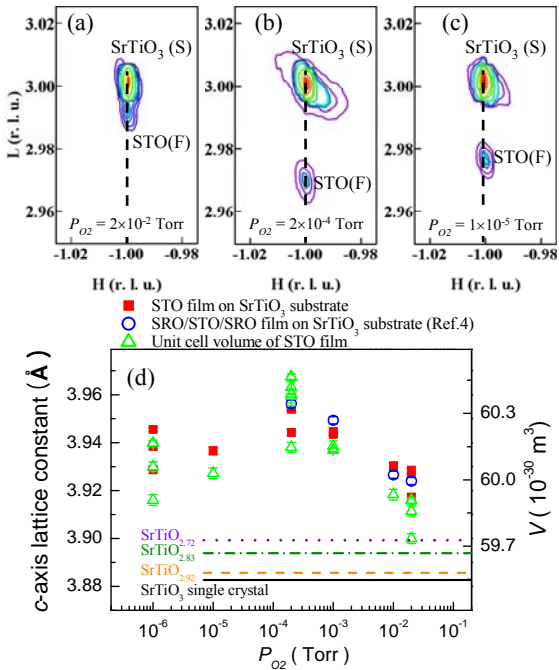

Fig. 1. (Color online) X-RSMs around the asymmetric (103) Bragg plane of STO films grown with oxygen partial pressure ($P_{O2}$) values of (a) $2\times10^{-2}$ Torr, (b) $2\times10^{-4}$ Torr, and (c) $1\times10^{-5}$ Torr. S and F denote the substrate and film, respectively. (1 reciprocal lattice unit (r.l.u.) = $2\pi/3.905$ Å$^{-1}$) (d) $P_{O2}$ dependence of structural properties of STO thin films. The solid (red) squares and open (green) triangles represent $c$-axis lattice constants and the unit cell volume, respectively. For comparison purposes, the $c$-axis lattice parameters of the SRO/STO/SRO heterostructure are denoted by the open (blue) circles from the work of Kim et al.[4].

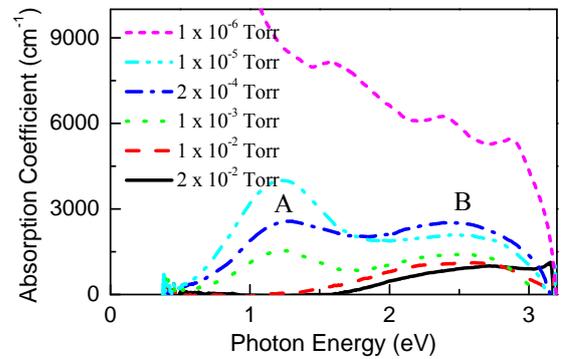

Fig. 2. (Color online) Visible and near-infrared absorption spectra of STO films grown at various values of $P_{O2}$, ranging from $2\times10^{-2}$ to $1\times10^{-6}$ Torr.

We investigated the structural properties of our STO thin films by using X-ray reciprocal space maps (X-RSMs). Figures 1(a), 1(b), and 1(c) show X-RSMs around the asymmetric ($\bar{1}03$) Bragg reflection of STO thin films grown at $P_{O2}$ values of $2\times10^{-3}$, $2\times10^{-4}$, and $1\times10^{-5}$ Torr, respectively. The X-RSMs show that the ($\bar{1}03$) Brag reflection peaks of the STO films and $SrTiO_3$ substrates were on the same ($H00$) line in reciprocal space. These results indicate that all of the STO films were perfect in-plane lattice matches for the $SrTiO_3$ substrates regardless of the $P_{O2}$ parameters.

These homoepitaxial STO films had almost the same lattice constants as those of our reported SRO/STO/SRO heterostructures, which exhibited room-temperature ferroelectricity.[4] The solid (red) squares in Fig. 1(d) show the $c$-axis lattice constants of the new STO films, deposited at various $P_{O2}$ values. This $P_{O2}$ dependence is nearly the same as that of the SRO/STO/SRO heterostructures,[4] which are denoted by open (blue) circles. Such an increase in tetragonality and expansion of the unit cell volume can be attributed to Coulomb repulsion between atoms around vacancy-type defects, which is easily introduced in oxide films during deposition.[5]

To obtain further insight, we investigated changes in the electronic structure by measuring optical absorption spectra. As shown in Fig. 2, we observed anomalous absorption peaks below the 3.2-eV optical band gap of bulk $SrTiO_3$, probably due to the formation of localized defect states in the STO films. In STO films grown at $P_{O2}$ values of $2\times10^{-2}$ and $1\times10^{-2}$ Torr, we observed a broad absorption peak centered at about 2.5 eV (absorption peak B). As $P_{O2}$ decreased further, we observed another absorption peak centered at about 1.3 eV (absorption peak A).

In our previous report,[4] we observed room-temperature ferroelectricity in STO heterostructures grown at $P_{O2}$ values between $1\times10^{-3}$ and $2\times10^{-4}$ Torr. In the present STO films grown under the same deposition conditions, absorption peak A was always present. Therefore, we suggest that absorption peak A and the associated changes in electronic structure are closely related to the room-temperature ferroelectricity in STO thin films.

We also performed X-ray absorption spectroscopy measurements of O 1s in the 7C1 beamline at the Pohang Light Source (results not shown here). We confirmed that there was no new state between the Fermi level and the conduction band bottom of the STO films. This suggests that the change in the electronic structure of FE STO films is located below the Fermi level.

To identify possible defects in the FE STO films, we investigated the stoichiometry using X-ray photoemission spectroscopy (XPS) and an electron-probe-micro-analysis (EPMA). From the ratio of the areas of Sr ($3d_{5/2}$ and $3d_{3/2}$) and Ti ($2p_{3/2}$ and $2p_{1/2}$) peaks in the XPS spectra, we determined that there were significant Sr deficiencies in our STO films. Similar EPMA results confirmed this. [Neither result is shown here.] Therefore, we decided to look into the formation of complex defects of O- and Sr-vacancies in our STO films.

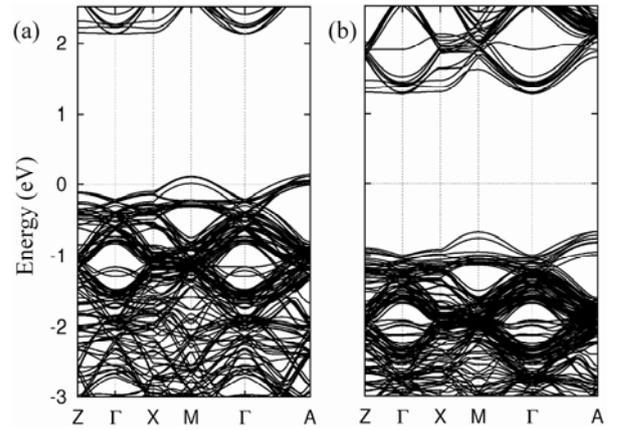

Fig. 3 (Color online) Results of first-principles calculations for the band structure of STO samples with (a) Sr vacancies and (b) Sr–O vacancies.

To determine the origins of the localized defect state that may be responsible for the optical absorption, we carried out density-functional-theory (DFT) calculations. We determined the equilibrium lattice structures of $SrTiO_3$ containing Sr, O, Sr–O, and Sr–O–O vacancy complexes in various configurations using a ($3\times3\times3$) supercell. The lattice constants of supercell units of $SrTiO_3$ that had vacancy complexes were optimized based on the experimentally observed $c/a$ ratio of 1.013. We performed DFT calculations using the local density approximation (LDA) and LDA+U for the electronic structures and atomic geometry optimization of $SrTiO_3$ with various vacancy configurations using the OpenMX code[6] based on the linear combination of localized pseudo atomic orbital method. We used Troullier-Martins-type norm-conserving pseudo potentials to replace the deep core potentials. For numerical integration and the solution of Poisson's equation, we used an energy cut-off of 300 Ry and performed the $k$-space integrations using a ($2\times2\times2$) $k$-grid. Since oxygen vacancy can provide an excess charge to the Ti-site resulting in significant $d$-orbital occupation, we employed the LDA+U calculations with the effective on-site Coulomb interaction parameter of $U_{eff}$ = 5 eV for Ti $3d$-states.[7,8] Our calculations predicted a direct band gap of 1.97 eV for pristine $SrTiO_3$, which is smaller than the experimental value of 3.2 eV but consistent with earlier theoretical calculation results.[9]

To begin with, we calculated the band structures of $SrTiO_3$ at various concentrations of point defects of Sr- or O-vacancies, such as $Sr_{0.875}TiO_3$, $Sr_{0.964}TiO_3$, $SrTiO_{0.875}$, and $SrTiO_{0.964}$. Similar to previous work,[10] the Fermi level of $SrTiO_3$ with oxygen vacancies was always in the conduction band (not shown). As shown in Fig. 3(a), the Fermi level of $SrTiO_3$ with Sr vacancies was in the valence band. These results imply that Sr or O vacancies both induce metallic $SrTiO_3$, unlike our insulating STO films. Therefore, point defects such as Sr or O vacancies can be excluded.

Charge-neutral Sr–O vacancy pairs can also be excluded from the possible defects in the STO. Although an Sr–O vacancy complex can make $Sr_{0.875}TiO_{2.875}$ and $Sr_{0.9634}TiO_{2.9634}$ retain their insulating qualities, the band calculations did not show any localized defect state in the band gap, as shown in Fig. 3(b). Therefore, the Sr–O

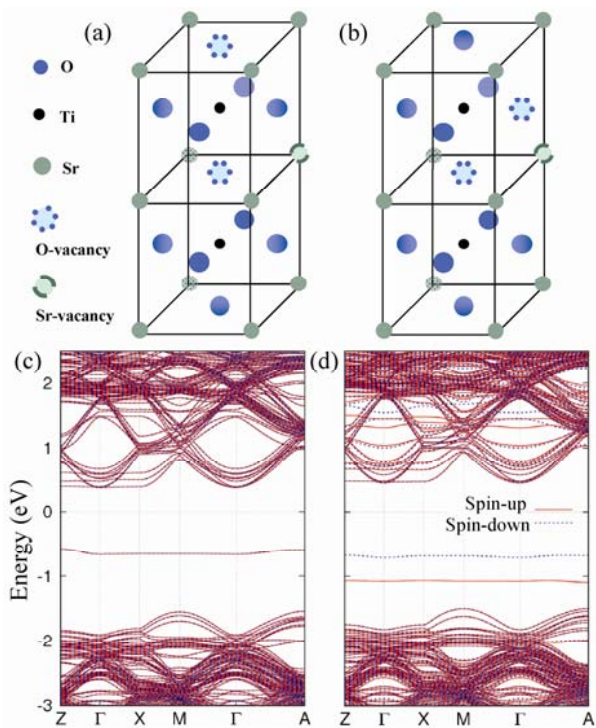

Fig. 4. (Color online) (a) and (c) are schematic configurations of Sr–O–O vacancies with the lowest total energies. (b) and (d) are calculated band structures for the configurations of (a) and (c), respectively.

vacancy complex cannot be responsible for peak A in Fig. 2.

As the next step in exploring possible defects in SrTiO$_3$, we considered the next-level vacancy complex consisting of Sr–O–O vacancy clusters. The total energy calculations of various configurations of Sr–O–O vacancy clusters showed that cluster formation is favored relative to the separate vacancies of the Sr, O, and Sr–O vacancy pair as $P_{O2}$ decreases. Figure 4 shows the two possible configurations of Sr–O–O vacancy clusters that have the highest values of binding energy and local dipole moments along the $c$-axis.

Figure 4(a) shows a configuration of Sr–O–O vacancy clusters in the tetragonal SrTiO$_3$ crystal structure where two oxygen vacancies are placed symmetrically around the Ti ion. As Fig. 4(c) shows, the corresponding band structure indicates that the insulating SrTiO$_3$ has localized and has doubly occupied the $e_g$ level below the Fermi level. Note that the energy difference between the $e_g$ state from the conduction band minimum is around 1 eV. When we consider the difference between the theoretical and the experimental band gap of pristine SrTiO$_3$, the optical transition from the localized defect state to the conduction band minimum can reasonably explain the observed absorption peak A located at 1.3 eV.

Figure 4(b) shows another Sr–O–O vacancy configuration, where both O vacancies are placed as close as possible to the Sr vacancy. As shown in Fig. 4(d), two electrons released from oxygen vacancies are transferred to the two nearby Ti atoms in the supercell, resulting in two singly occupied localized states. The localized state right above the valance band is from 3$d$-state of the Ti atom, which is surrounded by Sr–O–O vacancies and has an up-spin state. The down-spin state below the Fermi level is mainly from the $t_{2g}$ level of the Ti atom, located in the bottom right-hand unit cell. Note that the two states are located around 1.15 and 1.5 eV below the conduction band. Considering the broadness of absorption peak A, these localized states could also explain the experimentally observed absorption peak A.

Both of the above-mentioned Sr–O–O vacancies could explain the observed electronic structures of our FE STO films. However, the total energy of the configuration in Fig. 4(b) has a lower value than any of the other Sr–O–O vacancy configurations. The total energy of the configuration in Fig. 4(b) is 0.45 eV lower than that of the configuration in Fig. 4(c). This indicates that the Sr–O–O configuration in Fig 4(b) has the lowest formation energy. Therefore, the Sr–O–O vacancy configuration shown in Fig. 4(b) may be the most likely to be responsible for absorption peak A.

Since Sr–O–O vacancy complexes can form defect dipoles with their orientation determined by the vacancy configuration, they might result in ferroelectricity. However, it is still not clear how the orientation of Sr–O–O defect dipoles can be switched under an external electric field. Further investigation of the energetics of the dipole orientation changes is required.

In summary, we investigated changes in the electronic structure of strontium titanate thin films due to vacancy defects that could be created during deposition. By combining optical spectroscopy and first-principles calculations, we attributed the creation of the deep localized state that seems to have a close correlation with the appearance of ferroelectricity to the formation of Sr–O–O vacancy clusters. This interesting experimental finding can easily be used to develop advanced FE thin films or improve their properties.

This study was financially supported by Creative Research Initiatives (Functionally Integrated Oxide Heterostructures) of the Ministry of Science and Technology (MOST), the Korean Science and Engineering Foundation (KOSEF), and the KOSEF ARP (R17-2008-033-01000-0).